\documentclass[sigconf]{acmart}
\AtBeginDocument{%
  }

\copyrightyear{2026}
\acmYear{2026}
\setcopyright{cc}
\setcctype{by}
\acmConference[KDD '26]{Proceedings of the 32nd ACM SIGKDD Conference on Knowledge Discovery and Data Mining V.1}{August 09--13, 2026}{Jeju Island, Republic of Korea}
\acmBooktitle{Proceedings of the 32nd ACM SIGKDD Conference on Knowledge Discovery and Data Mining V.1 (KDD '26), August 09--13, 2026, Jeju Island, Republic of Korea}
\acmDOI{10.1145/3770854.3780324}
\acmISBN{979-8-4007-2258-5/2026/08}

\usepackage{makecell}
\usepackage{xcolor}
\usepackage{colortbl}
\usepackage{xspace}
\usepackage{enumitem}
\usepackage{multirow}

\newcommand{\eg}{\textit{e.g.},\ }
\newcommand{\ie}{\textit{i.e.},\ }
\newcommand{\techname}{SPiKE}

\newcommand{\ours}[1]{\cellcolor{gray!15}{\textbf{#1}}}
\newcommand{\first}[1]{\cellcolor{gray!15}{\textbf{#1}}}

\begin{document}

\title{Enriching Semantic Profiles into Knowledge Graph for Recommender Systems Using Large Language Models}

\author{Seokho Ahn}
\orcid{0000-0002-5715-4057}
\affiliation{%
  \institution{Inha University}
  \department{Department of Electrical and Computer Engineering}
  \city{Incheon}
  \country{South Korea}}
\email{sokho0514@inha.edu}

\author{Sungbok Shin}
\orcid{0000-0001-6777-8843}
\authornotemark[1]
\affiliation{%
  \institution{Sogang University}
  \department{Department of Computer Science \\and Engineering}
  \city{Seoul}
  \country{South Korea}}
\email{sbshin90@sogang.ac.kr}

\author{Young-Duk Seo}
\orcid{0000-0001-8542-2058}
\authornote{Co-corresponding authors.}
\affiliation{%
  \institution{Inha University}
  \department{Department of Electrical and Computer Engineering}
  \city{Incheon}
  \country{South Korea}
}
\email{mysid88@inha.ac.kr}






\renewcommand{\shortauthors}{Seokho Ahn, Sungbok Shin, and Young-Duk Seo}

\begin{abstract}


Rich and informative profiling to capture user preferences is essential for improving recommendation quality.
However, there is still no consensus on how best to construct and utilize such profiles.
To address this, we revisit recent profiling-based approaches in recommender systems along four dimensions: 1) knowledge base, 2) preference indicator, 3) impact range, and 4) subject.
We argue that large language models (LLMs) are effective at extracting compressed rationales from diverse knowledge sources, while knowledge graphs (KGs) are better suited for propagating these profiles to extend their reach.
Building on this insight, we propose a new recommendation model, called \textbf{SPiKE}. 
SPiKE consists of three core components:
i) Entity profile generation, which uses LLMs to generate semantic profiles for all KG entities;
ii) Profile-aware KG aggregation, which integrates these profiles into the KG; and
iii) Pairwise profile preference matching, which aligns LLM- and KG-based representations during training.
In experiments, we demonstrate that SPiKE consistently outperforms state-of-the-art KG- and LLM-based recommenders in real-world settings.
\end{abstract}


%
%
\maketitle
\newcommand\kddavailabilityurl{https://doi.org/10.5281/zenodo.18041669}
\ifdefempty{\kddavailabilityurl}{}{
\addvspace{-0.7em}
\begingroup\small\noindent\raggedright\textbf{Resource Availability:}\\
The implementation code of this paper has been made publicly available at \url{\kddavailabilityurl}.
\endgroup
}

\section{Introduction}

Obtaining large volumes of data is essential for imitating human-like systems with high performance, but acquiring personal data at scale is difficult~\cite{birch2021data}. 
This is also true for recommender systems~\cite{mekouar2024global}. 
Rich and extensive user profiles are crucial for producing accurate personalized recommendations, whether for films~\cite{lavanya2021comprehensive, zhang2024generative,wang2025lettingo}, retail products~\cite{ma2025pub, wang2025lettingo, fan2023adversarial}, or other items~\cite{xiao2023uprec, schedl2022lfm}. 
However, profile information is often incomplete or unavailable, and thus can potentially compromise recommendation accuracy~\cite{ren2024rlmrec}.

Recent prior work proposes various construction methodologies for enriching profiles and propagating them using knowledge graphs (KGs) \cite{wang2019kgat, wang2021kgin, yang2023kgrec, yang2022kgcl, bufi2024kguf}, large language models (LLMs) \cite{wu2024rssurveyllms, ren2024rlmrec, xi2023kar, gao2023chatrec, zhu2024cllm4rec, wei2024llmrec}, or their combination \cite{cui2025colakg}. 
However, the optimal \textit{modus operandi} for constructing user profiles and propagating them in recommender systems remains unclear.
To this end, we aim to address this knowledge gap by identifying the most desirable method for enriching profiles on recommender systems. 
We begin by evaluating the choices from prior work across the following four dimensions: 1) knowledge base, 2) preference indicator, 3) impact range, and 4) subject, then identify the profiling method that most closely aligns with our criteria. 
Afterwards, we select the model that best accommodates our optimal choice of profiling method.

Based on our optimized approach, we develop a recommendation model called \textbf{\techname{}} (\textbf{S}emantic \textbf{P}rofiles \textbf{i}nto \textbf{K}nowledge Graphs for \textbf{E}nhanced Recommendation).
SPiKE improves recommendation performance by enriching KG with semantic profiles that effectively capture user preferences. 
Rather than relying on LLMs to interpret KG structures, profiles are designed for structural incorporation and propagation in KGs.
To this end, SPiKE consists of three key components:
i) \textbf{Entity profile generation} (\S\ref{sec:entity_profiling}) constructs high-quality textual profiles for all KG entities; 
ii) \textbf{Profile-aware KG aggregation} (\S\ref{sec:kg_aggregation}) incorporates these profiles into the KG to jointly propagate structural and semantic signals; and
iii) \textbf{Pairwise profile preference matching} (\S\ref{sec:pairwise_matching}) encourages similar profiles to be embedded close together, and dissimilar ones to be placed farther apart.
In summary, SPiKE follows four design principles, prioritizing simplicity and core functionality to highlight the complementary roles of LLM-based profiling and KG-based propagation.

Comprehensive results show that SPiKE consistently outperforms state-of-the-art recommendation models including knowledge aware, LLM-based, and hybrid approaches, across three real world benchmarks, highlighting the benefits of our design choices.

\begin{table*}[t!]
\centering
\caption{A Comparison of profiling methods across different recommender systems (RS) models between 2021 and 2025. 
We taxonomize 10 recent prior works into 4 dimensions: 1) knowledge base, 2) preference indicator, 3) impact range, and 4) subject. 
Based on this taxonomy, we identify the desirable profiling method (\S\ref{subsec:approach-profiles}) and select the model that best suits the task (\S\ref{subsec:related-llm}).
}
\vspace{-0.7em}
\setlength\tabcolsep{4pt}
\resizebox{1\textwidth}{!}{%
\begin{tabular}{llllll}
\toprule
\textbf{RS model} & \textbf{Works} & \textbf{Knowledge base} & \textbf{Preference indicator} & \textbf{Impact range} & \textbf{Subject} \\
\cmidrule(l{0.5mm}r{0.5mm}){1-2} \cmidrule(l{0.5mm}r{0.5mm}){3-6}
\textbf{KG}      & \cite{wang2021kgin, yang2023kgrec, yang2022kgcl, bufi2024kguf}        & User feedback, KG & Entity connection  & Global KG & User-item interaction \\
\cmidrule(l{0.5mm}r{0.5mm}){1-2} \cmidrule(l{0.5mm}r{0.5mm}){3-6}
\multirow{2}{*}{\textbf{LLM}}     & \cite{xi2023kar, gao2023chatrec, zhu2024cllm4rec, liu2023once}             & User feedback, textual metadata, LLM prior & Item description* & Profile itself & User profile \\
                 & \cite{ren2024rlmrec}             & User feedback, textual metadata, LLM prior & Item profile* & Profile itself & User/Item profiles \\
\cmidrule(l{0.5mm}r{0.5mm}){1-2} \cmidrule(l{0.5mm}r{0.5mm}){3-6}
\multirow{2}{*}{\textbf{KG+LLM}}  & \cite{cui2025colakg}             & User feedback, KG, LLM prior & KG subset* & Local KG & User/Item profiles \\
& \ours{\textbf{SPiKE (Ours)}} & \ours{User feedback, textual metadata, KG, LLM prior}  & \ours{KG subset, item profile*} & \ours{\textbf{Global KG}} & \ours{Entity profiles} \\
\bottomrule
\end{tabular}
}
\\[0.3em]
\rightline{\small{\;\;* These indicators are provided as prompt-level inputs to instruct the LLM.}\;}
\vspace{-0.8em}
\label{tb:dimension}
\end{table*}

In summary, the contributions of our work are: 
\begin{itemize}[noitemsep, leftmargin=1.1em]
\item We identify four key dimensions that guide the construction and utilization of enriched profiles for improved recommendations.
\item We propose a recommender model SPiKE, designed to address these dimensions using our optimal profiling strategy.
\item Experiments on real-world benchmarks show that SPiKE consistently outperforms non-profiling and profiling-based methods.
\end{itemize}

\section{Overview \label{sec:02_overview}}
This section provides an overview of our approach. 
After a brief introduction on profiling (\S\ref{subsec:profiliing-rec}), we present our dimensions for enriching profiles, and the rationales for developing the model (\S\ref{subsec:approach-profiles}). 
We conclude with a discussion of our design choices (\S\ref{subsec:related-llm}). 

\subsection{Profiling in Recommendation}
\label{subsec:profiliing-rec}
We begin this section by defining and clarifying the scope of the term \textit{profiling} in recommender systems in our context.
Profiling is the automated process of constructing a structured representation of an entity by analyzing the information associated with it \cite{wang2019userprofiling,yan21profiling,gu20hierrecsys}.
The resulting representation, which we call a \textit{profile}, infers and encodes the salient attributes of that entity.
Traditionally, profiling has focused on users, summarizing their interests, traits, and behavioral patterns \cite{eke2019survey,raza2025reviewrs, purificato2024usermodeling,tan2023usermodelingeralarge}.
However, due to the private nature of user-side information \cite{mekouar2024global}, recent approaches have explored non-user signals (\eg item description and category) that implicitly reflect user preferences \cite{xi2023kar,ren2024rlmrec,cui2025colakg,raza2025reviewrs,bufi2024kguf,wu2024rssurveyllms}.
In particular, several works generate item profiles that summarize not only the item-side attributes but also user preferences inferred from these signals, effectively capturing preferences beyond the user profile~\cite{ren2024rlmrec,cui2025colakg,wu2024rssurveyllms}. 
From this broader perspective, profiling in recommender systems can refer to any process, such as user- or item-based, that captures user preferences.

Profiles in recommender systems take various forms, with recent work largely following two directions: 
(i) textual profiles generated by large language models (LLMs) \cite{wu2024rssurveyllms, ren2024rlmrec, cui2025colakg, xi2023kar, gao2023chatrec, zhu2024cllm4rec, wei2024llmrec} which summarize rich and diverse textual information into concise paragraphs that explain user preferences; and 
(ii) structured profiles derived from knowledge graphs (KGs) \cite{wang2019kgat, wang2021kgin, yang2023kgrec, yang2022kgcl, bufi2024kguf, cui2025colakg}, which implicitly capture preferences by propagating structural signals over multi-hop relationships.

While both capture key characteristics of users and items, their structural differences lead each approach to emphasize different aspects of user preference.
For example, a 1-hop neighborhood in KG reveals basic factual signals, such as a user's items or their attributes that they interacted in the past (\eg descriptions and categories).
Multi-hop neighbors provide richer context than 1-hop neighborhood (\eg user-user or item-item connections) but often result in unstructured large subgraphs that are hard to understand \cite{cui2025colakg}.
In contrast, LLMs can summarize item-side information and infer missing knowledge, but do not have user-side information, unless it is prompted by the user \cite{mekouar2024global, arora2022public}.

\subsection{Dimensions for Enriching Profiles}
\label{subsec:approach-profiles}
While profile plays an important role in recommendation, how it should be constructed and applied remains unclear.
To this end, we investigate how user preference-aware profiles can be constructed and which system architectures best support them.
Specifically, we examined existing works along four dimensions that together show how accurate and effective a recommender system can capture user preferences: \textbf{1) Knowledge base}, \textbf{2) Preference indicator}, \textbf{3) Impact range}, and \textbf{4) Subject}.
We then provide a four-dimensional comparison of existing methods (\ie KG-based methods, LLM-based methods, and combined methods) shown in Table \ref{tb:dimension}.
Detailed descriptions of each dimension are provided below:

\begin{enumerate}[itemsep=0.1cm, leftmargin=-0.0cm, itemindent=0.5cm,label=\textbf{\arabic *)}]

\item \textbf{Knowledge base} refers to the \textit{richness} of data sources for constructing profiles. 
A richer base enables more precise modeling of user preferences by capturing both user- and item-side characteristics.  
For example, instead of stating that a user likes a specific movie (\eg user feedback), a richer base enables reasoning that the user prefers movies characterized by certain attributes such as genre, director, or audience type. 
KGs provide structured connections between entities. 
Pretrained LLMs offer item-side factual priors without additional tuning, and textual metadata (\eg item descriptions) captures content-level preferences.
However, excessive information can introduce noise, making filtering and summarization necessary \cite{ren2024rlmrec}. 
This poses a challenge for KG-based approaches, but LLM-based models handle this issue more effectively.

\item \textbf{Preference indicator} works as the \textit{evidence} used to indicate user preference.  
While the knowledge base supplies background information, the preference indicator retains evidential user preference that guides the model's decisions.
For example, KG-based models infer user preferences from structural connections between nodes, whereas LLMs derive them from the prompt’s explicit content.
When using LLMs, the challenge lies in adding high-quality information, despite the high computational cost and limited context windows.
The key to addressing this is in either effective selection \cite{xi2023kar, gao2023chatrec, zhu2024cllm4rec, liu2023once}, or compression of preference contents \cite{ren2024rlmrec}.

\item \textbf{Impact range} refers to how far a profile's influence \textit{extends} in the recommendation model.  
As noted above, profiles generated by pretrained LLMs remain in themselves. 
KG‑based methods implicitly propagate profile information through aggregation, achieving broader reach even without explicit alignment. 
However, because they solely rely on the KG, they cannot incorporate richer external sources.
When the impact range is limited, sparsely connected users and items receive little influence from the profile. 
This eventually leads to degraded recommendation performance \cite{zou2022improving}.
To counter this challenge, hybrid systems that utilize both large LLMs with KGs have emerged \cite{cui2025colakg}. 
Still, these methods confine profile signals to local user-item modeling, instead of propagating them throughout the entire KG. 
To fully take advantage of profiles that contain accurate user preferences, it is crucial to design mechanisms that extend their influence to a wider range.

\item \textbf{Subject} denotes the \textit{target} entity for which the model explicitly captures a user's preferences in the profile. 
Explicit user preference descriptions reveal what the user pays attention to, enabling accurate recommendations.
KG-only models provide explicit links between users and items, but the specific attributes that users prefer remain implicit.
In contrast, LLM-generated user profiles are explicit, but they are generally focused on the user, often limited to item coverage.
Capturing attribute-level preferences demands profiling entities beyond just users and items. 
Recent work \cite{ren2024rlmrec, cui2025colakg} moves in this direction by adding item profiles (\eg ``this item appeals to users who value attribute A''). 
Nonetheless, coverage remains confined to the user-item level, leaving the model unable to accurately express which specific item attributes are preferred.

\end{enumerate}

\paragraph{Our design rationale.}
Based on the four profiling dimensions, we propose four design rationales for the model:

\begin{enumerate}[noitemsep, leftmargin=0.6cm]
\item [\textbf{R1}] \textbf{Accumulating knowledge. } We collect various sources of knowledge (\eg structured feedback, KG relations, item metadata, and user reviews), so that the model
can take advantage of various reasonings. 
\item [\textbf{R2}] \textbf{Deepening preference.} We detail and improve the accuracy of the preference information through accumulating information and refinement (\eg reasoning and summarizing).
\item [\textbf{R3}] \textbf{Broadening propagation.} We aim to propagate profiles to a wider set of entities.
\item [\textbf{R4}] \textbf{Expanding subjects.} We extend profiling beyond users and items to all entities to enrich the alignment between item characteristics and user tastes.
\end{enumerate}

\subsection{Our Approach for Model Selection\label{subsec:related-llm}}
Recent research on LLMs and knowledge-aware recommendation can be broadly categorized into LLM-only, KG-only, and hybrid approaches. 
LLM-based methods contribute in two primary ways: 
either as recommenders that directly interpret user profiles to generate recommendations \cite{wu2024rssurveyllms, tan2023usermodelingeralarge, zhao2025surveyllms, wang2023zeroshot, bao2023tallrec, lin2024rella, Wang2025_re2llm, li2023pmmrec}, 
or as profilers that extract and restructure item-side information (\eg descriptions and reviews) into semantically rich representations \cite{wu2024rssurveyllms, tan2023usermodelingeralarge}. 
The former typically demands extensive parameter tuning and struggles with scalability, while the latter (\ie prompted without fine-tuning) leverages LLMs as frozen backbones for summarization \cite{ren2024rlmrec, cui2025colakg, xi2023kar, gao2023chatrec, zhu2024cllm4rec, wei2024llmrec}, semantic enrichment \cite{cui2025colakg, ren2024rlmrec, liu2023once}, and knowledge inference \cite{cui2025colakg, wei2024llmrec}. 
However, these LLM-based profiling approaches are inherently constrained by limited access to user-side information and prompt length restrictions, resulting in narrow contextual coverage.

On the other hand, knowledge-aware recommenders derive profile-like signals from structured data such as user-item interactions and multi-hop paths within a KG. 
Path-based models \cite{wang2019kprn, wang2022remr, xian2019pgpr, karidi2025path, geng2022plmrec} offer interpretability but often rely on brittle heuristics \cite{zhu2024kpcl, cui2025colakg}, whereas aggregation-based methods \cite{wang2019kgat, wang2021kgin, yang2023kgrec, yang2022kgcl, bufi2024kguf, cui2025colakg} achieve higher accuracy by capturing implicit, high-order relations. 
Despite their strengths, these KG-based methods remain bounded by the incompleteness of predefined knowledge graphs, which can hinder recommendation quality.

To address the limitations of LLM-only and KG-only paradigms, we adopt a hybrid strategy that leverages the strengths of each: LLMs are employed as profilers to construct semantically enriched item-side representations, while KGs propagate these profiles throughout the graph. 
This design mitigates prompt-length constraints by offloading propagation to the KG, effectively scaling the influence of LLM-generated profiles beyond their local context.

We implement this strategy in \textbf{SPiKE} (\textbf{S}emantic \textbf{P}rofiles \textbf{i}nto \textbf{K}nowledge Graphs for \textbf{E}nhanced Recommendation), a model built on a hybrid knowledge base that combines structured user-item interactions and KG relations with unstructured textual metadata (\eg item descriptions and user reviews), thereby incorporating a wide range of information sources \textbf{(R1)}. 
In SPiKE, LLMs receive input evidence from this knowledge base and are prompted with item-side information to enrich user profiles through summarization, reasoning, and noise reduction \textbf{(R2)}. 
Following prior work \cite{ren2024rlmrec, cui2025colakg}, item profiles are injected into prompts to compensate for missing user-side signals. The resulting profiles are propagated across the KG using message-passing aggregation, allowing preference signals to influence distant nodes \textbf{(R3)}. 
Furthermore, SPiKE generalizes profiling to all KG entities, not just users and items, thereby enabling reasoning over auxiliary semantic content and improving the alignment between item characteristics and user preferences \textbf{(R4)}. 
We argue that enabling KGs to interpret and propagate LLM-generated profiles is more effective and scalable than requiring LLMs to reason directly over KG structures.

\begin{figure*}
    \centering
    \includegraphics[width=0.98\textwidth]{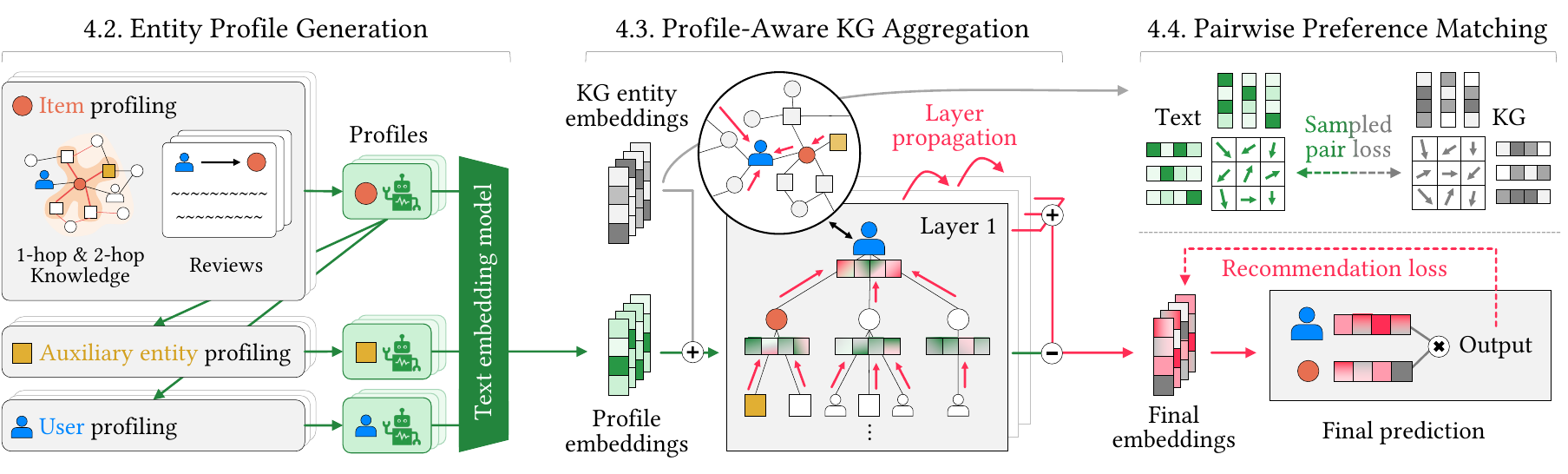}
    \vspace{-0.7em}
    \caption{Illustrative flow of \textbf{SPiKE}. SPiKE is built on the idea from \S\ref{sec:02_overview} that LLMs and KGs each contribute their essential strengths (\ie semantic profiling and structural propagation). Following a minimal-role design, SPiKE generates entity profiles using LLMs (\S\ref{sec:entity_profiling}), integrates them into KG aggregation (\S\ref{sec:kg_aggregation}), and applies pairwise matching (\S\ref{sec:pairwise_matching}) during training.}
    \Description[]{Illustrative flow of \textbf{SPiKE}. SPiKE is built on the idea from \S\ref{sec:02_overview} that LLMs and KGs each contribute their essential strengths (\ie semantic profiling and structural propagation). Following a minimal-role design, SPiKE generates entity profiles using LLMs (\S\ref{sec:entity_profiling}), integrates them into KG aggregation (\S\ref{sec:kg_aggregation}), and applies pairwise matching (\S\ref{sec:pairwise_matching}) during training.}
    \label{fig:spike}
    \vspace{0.5em}
\end{figure*}

\section{Task Definition}

Building on the above design considerations, this section formalizes the problem setting by introducing the key notations for the knowledge graph (KG) and the profile-enriched KG (PKG), and defining our PKG-aware recommendation task.

\subsubsection*{Knowledge graph in recommender systems}
A \textit{knowledge graph (KG)} is a directed graph of nodes and edges that provides auxiliary knowledge about items \cite{yang2023kgrec, wang2019kgat}.
We formally define a KG as:
\begin{equation}
    \mathcal{G} = \left\{(h, r, t) \mid h, t \in \mathcal{E},\ r \in \mathcal{R} \right\},
\end{equation}
where \(\mathcal{E}\) denotes the set of entities and \(\mathcal{R}\) is a set of relation types that describe how entities are semantically connected. 
In recommender systems, we assume that users \(\mathcal{U}\) and items \(\mathcal{V}\) are subsets of entities \(\mathcal{E}\), \ie \(\mathcal{U}\cup\mathcal{V} \subseteq \mathcal{E}\). 
Users and items are linked by the relation \(r_\mathsf{Interact} \in \mathcal{R}\) when a user \(u \in \mathcal{U}\) interacts with an item \(v \in \mathcal{V}\), \ie \((u, r_\mathsf{Interact}, v) \in \mathcal{G}\).
Auxiliary entities other than users \(\mathcal{U}\) and items \(\mathcal{V}\) serve as additional information related to them. 
For instance, in the book recommendation domain, such entities are connected to items via triples like \((v_\mathsf{Wicked}, r_\mathsf{Genre}, e_\mathsf{Fantasy}) \in \mathcal{G}\).

\subsubsection*{Profile-enriched knowledge graph}
We first define the concept of a \textit{profile-enriched KG (PKG)} as an extension of a standard KG, where each entity is augmented with additional semantic knowledge.  
Specifically, each entity \(e \in \mathcal{E}\) is paired with a \textit{profile} \(p_e\), a text paragraph that summarizes key attributes of the entity and captures preferences to support high-quality personalized recommendation.
Using entity profiles, we formally define the PKG as
\(\overline{\mathcal{G}}=\left<\mathcal{G}, \mathcal{P}\right>\), where \(\mathcal{P} = \left\{p_e \mid e \in\mathcal{E}\right\}\) denotes the set of profiles.
We further explain how profiles are constructed and what textual components should be included in \S \ref{sec:entity_profiling}.

\subsubsection*{Task formulation}
We formulate our task as a PKG-aware recommendation problem, where the objective is to predict user preferences by leveraging PKG \(\overline{\mathcal{G}}=\left<\mathcal{G}, \mathcal{P}\right>\) constructed for training.
The recommender model \(\mathcal{F}(u, v | \Theta_{\textrm{opt}}, \overline{\mathcal{G}})\) aims to estimate the preference score \(\hat{y}_{uv} \in \left[0, 1\right]\) for any user \(u \in \mathcal{U}\) and item \(v \in \mathcal{V}\).
The model parameters \(\Theta_{\textrm{opt}}\) are optimized by minimizing a loss function \(\mathcal{L}\).

\section{Methodology}
This section outlines our method, \textbf{SPiKE} (\textbf{S}emantic \textbf{P}rofiles \textbf{i}nto \textbf{K}nowledge Graphs for \textbf{E}nhanced Recommendation).

\subsection{SPiKE Overview}
Figure \ref{fig:spike} illustrates the overview of SPiKE. To enhance recommendation quality, SPiKE enriches the KG with semantic profiles that capture user preferences. 
The output of SPiKE is a recommendation score \(\hat{y}_{uv}\) for a given user-item pair \((u, v)\) through the following three components: 
(i) \textbf{Entity profile generation} (\S\ref{sec:entity_profiling}) constructs high-quality textual profiles for all entities, reflecting user preferences; 
(ii) \textbf{Profile-aware KG aggregation} (\S\ref{sec:kg_aggregation}) incorporates these profiles into the KG to jointly propagate structural and semantic signals; and
(iii) \textbf{Pairwise profile preference matching} (\S\ref{sec:pairwise_matching}) encourages similar profiles to be embedded close together, and dissimilar ones farther apart. 
SPiKE follows the design principles in \S \ref{subsec:approach-profiles}, prioritizing simplicity and core functionality to focus on the complementary roles of LLM-based profiling and KG-based propagation.
The remainder of this section details each component.

\begin{figure*}
    \centering
    \includegraphics[width=0.98\textwidth]{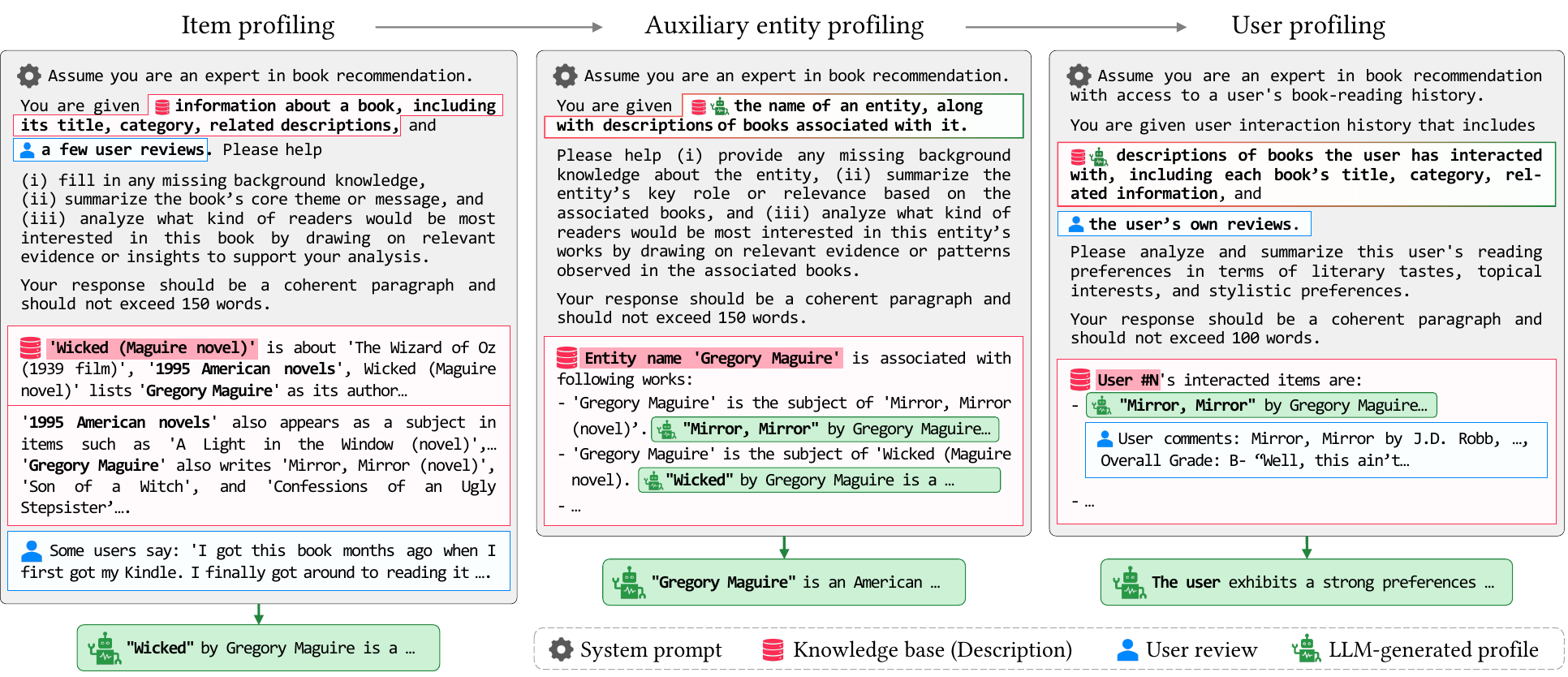}
    \vspace{-0.7em}
    \caption{Examples of entity profile generation prompts used in SPiKE for book recommendation.
    Each entity in the KG is profiled in the order of item, auxiliary entity, and user, incorporating relevant knowledge bases and preference indicators.
    }
    \Description[]{Examples of entity profile generation prompts used in SPiKE for book recommendation.
    Each entity in the KG is profiled in the order of item, auxiliary entity, and user, incorporating relevant knowledge bases and preference indicators.
    }
    \vspace{-0.75em}
    \label{fig:profile_prompt}
\end{figure*}

\subsection{Entity Profile Generation Using LLMs}
\label{sec:entity_profiling}

The goal of this section is to generate semantic profiles \(\mathcal{P}\) for each entity \(e \in\mathcal{E}\). 
Well-constructed profiles should reflect user-oriented preferences, indicating which users are likely to prefer a given entity and why \cite{ren2024rlmrec}.
To achieve this, we leverage LLMs to complement factual information and summarize preference-related aspects \cite{ren2024rlmrec, cui2025colakg}.
%
Unlike prior works \cite{ren2024rlmrec, cui2025colakg, xi2023kar, gao2023chatrec, zhu2024cllm4rec, wei2024llmrec}, we highlight that profiles are constructed for the entire entity set \(\mathcal{E}\), including users, items, and auxiliary entities.

The profiling pipeline proceeds bottom‑up \cite{ren2024rlmrec}: it first constructs item profiles, then auxiliary-entity profiles, and finally user profiles. 
Illustrative example prompts are shown in Figure~\ref{fig:profile_prompt}.
Public knowledge about items and auxiliary entities (\eg genre, actor, or director in the movie domain) is well-documented and can be readily summarized by LLMs \cite{arora2022public}, whereas user information is typically unavailable to KGs and LLMs due to its private nature \cite{mekouar2024global, arora2022public}.
Thus, constructing item and auxiliary profiles first supplies a rich semantic context from which user profiles can be inferred.
We now describe the profile construction process for each type in detail.

\subsubsection*{Item profiling}
To construct \(\mathcal{P}_{\mathsf{Item}}\), each item profile \(p_v \in \mathcal{P}_\mathsf{Item}\) is generated by prompting the LLM to supplement attribute knowledge, summarize the content, and infer user preferences along with their rationale.  
Specifically, we combine the system prompt \(I_v\), factual descriptions \(D_v\) from the knowledge graph (including 1-hop and 2-hop entity attribute), and user reviews \(R_v\) that reflect individual preferences for item \(v \in \mathcal{V}\):
\begin{equation}
    p_v = \mathsf{LLM}\left(I_v, D_v, R_v\right),\quad
    \mathcal{P}_{\mathsf{Item}} = \left\{p_v \mid v \in \mathcal{V} \right\}
\end{equation}
To construct \(D_v\), each 1-hop triplet is converted into a predefined textual template (e.g., \(\{v_{\textsf{Wicked}}\}\) \textsf{belongs to literary genre} \(\{e_{\textsf{Fantasy}}\}\)) for better interpretability by the LLM.  
To incorporate 2-hop knowledge, we follow \citet{cui2025colakg} to sample items from 1-hop entities. 
These items are then converted into a template (e.g., \(\{e_{\textsf{Fantasy}}\}\) \textsf{also appears as a genre of items such as} \(\{v_1, v_2, \ldots\}\)).  
Reviews in \(R_v\) are randomly sampled due to prompt length constraints.
Please refer to Appendix \ref{sec:appendix_template} for textual template examples.

\subsubsection*{Auxiliary entity profiling}
To construct auxiliary entity profiles \(\mathcal{P}_{\mathsf{Aug}}\), we leverage existing item profiles as informative input to enable high-quality descriptions.
Each entity \(e \in \mathcal{E} - (\mathcal{U} \cup \mathcal{V})\) is summarized by the LLM using a system prompt \(I_e\) and a subset of related item profiles, \ie \(P_e \subseteq \{p_h \in \mathcal{P}_\mathsf{Item} \mid (h, r, t) \in \mathcal{G},\ t = e\}\):
\begin{equation}
    p_e = \mathsf{LLM}\left(I_e, P_e\right),\quad
    \mathcal{P}_{\mathsf{Aug}} = \left\{p_e \mid e \in \mathcal{E} - (\mathcal{U} \cup \mathcal{V}) \right\}.
\end{equation}
Similar to item profiling, LLM is guided to describe the entity, infer user preferences, and provide reasoning.

\subsubsection*{User profiling}
Unlike items or entities, user information is inherently private and often limited.  
To address this, we leverage the previously constructed item profiles to enrich a user's interaction history:  
\(P_u \subseteq \{p_t \in \mathcal{P}_\mathsf{Item} \mid (h, r_{\mathsf{Interact}}, t) \in \mathcal{G},\ h = u\}\),  
by providing informative descriptions.  
By appending the user's reviews \(R_u\) to the item profiles in \(P_u\), we obtain a richer context for profiling.
The LLM is then prompted to summarize the user's preferences from both item and entity perspectives:
\begin{equation}
    p_u = \mathsf{LLM}\left(I_u, P_u, R_u\right),\quad
    \mathcal{P}_{\mathsf{User}} = \left\{p_u \mid u \in \mathcal{U} \right\}.
\end{equation}

\subsubsection*{Entity profile representation}
After applying different profiling strategies, the final semantic profiles \(\mathcal{P}\) are formed by combining the three subsets, \(\mathcal{P}=\mathcal{P}_{\mathsf{Item}}\cup\mathcal{P}_{\mathsf{Aug}}\cup\mathcal{P}_{\mathsf{User}}\).
Each profile \(p_e \in\mathcal{P}\) is then embedded into \(d_s\)-dimensional vector representation \(\mathbf{p}_e = \mathcal{F}_{\mathsf{Text}}(p_e)\) via pre-trained text encoder \(\mathcal{F}_{\mathsf{Text}}:\mathcal{P}\rightarrow \mathbb{R}^{d_s}\), ensuring alignment with the profile-enriched KG.

\subsection{Profile-Aware KG Aggregation}
\label{sec:kg_aggregation}

This section describes how the generated semantic profiles are integrated into the knowledge graph for representation learning.
To do so, we first inject each profile as an additive semantic signal during neighborhood aggregation.
After message passing, the profile is removed to ensure that the learned embeddings remain in the original KG space.
This ResNet \cite{he2015resnet}-like coupling with opposite sign lets the aggregation ``touch'' the semantic profile; an intuitive way to inject semantic information when weighing neighbors, while preserving the original KG structure.

Formally, let \(\mathbf{x}_e\in\mathbb{R}^d\) denote the \(d\)-dimensional initial entity embedding \(e \in \mathcal{E}\), and let \(\mathbf{p}_e\in\mathcal{P}\) be its profile embedding obtained in the previous section.
We form a profile-aware representation as:
\begin{equation}
\overline{\mathbf{x}}_e = \mathbf{x}_e + \lambda_{p}\cdot\mathcal{M}(\mathbf{p}_e)
\end{equation}
where \(\mathcal{M}:\mathbb{R}^{d_s}\rightarrow \mathbb{R}^{d}\) denotes a two-layer perceptron, a simple but effective choice \cite{ren2024rlmrec, cui2025colakg}, that projects the profile embedding \(\mathbf{p}_e\) into the recommendation space.
\(\lambda_{p}\) is a scaling factor that controls its influence.
Then, given a triplet \((h,r,t)\in\mathcal{G}\), we adapt the score weighting function \(\phi(h, r, t)\) from \cite{yang2023kgrec, zhu2024kpcl} to compute a profile-aware neighborhood score, conditioning it on the profile-augmented head \(h\), relation \(r\), and tail \(t\):
\begin{equation}
\phi(h, r, t) =
\mathsf{Softmax}_{\mathcal{N}_h} \left[
\frac{
\overline{\mathbf{x}}_h \mathcal{W}_1 \cdot 
\left( \overline{\mathbf{x}}_{t} \mathcal{W}_2 \odot \overline{\mathbf{x}}_{r} \right)^\top
}{
\sqrt{d}
}
\right],
\label{eq:profile-score}
\end{equation}
where \(\mathcal{W}_1, \mathcal{W}_2\in \mathbb{R}^{d \times d}\) are learnable attention  parameters, \(\odot\) denotes the element-wise product, and \(\mathsf{Softmax}_{\mathcal{N}_h}\) is the softmax operator applied over the set of neighbors \(\mathcal{N}_h=\left\{(h', r, t)\in\mathcal{G} \mid h=h'\right\}\) of entity \(h\in\mathcal{E}\).

The next step is to aggregate profile-aware messages using the computed neighborhood scores. 
For non-user heads, relation and tail signals are jointly used.
For user heads, we use only neighbor embeddings to handle privacy-limited information.
This design choice aligns with earlier KG-based approaches \cite{yang2023kgrec, bufi2024kguf}.
The aggregation at the \(l\)-th layer is defined as:
\begin{equation}
\overline{\mathbf{x}}_h^{(l)} =
\begin{cases}
\textsf{Agg}_{\mathcal{N}_h}\left[
\overline{\mathbf{x}}_t^{(l-1)}
\right], & \textrm{if } h \in \mathcal{U},
 \\[10pt]
\textsf{Agg}_{\mathcal{N}_h}\left[
\phi(h,r,t) \cdot \left( \overline{\mathbf{x}}_r \odot \overline{\mathbf{x}}_t^{(l-1)} \right)
\right]
, & \textrm{otherwise.}
\end{cases}
\label{eq:profile-agg}
\end{equation}
Here, \(\textsf{Agg}_{\mathcal{N}_h}\) denotes a mean aggregator over the neighborhood \(\mathcal{N}_h\), defined as \(\mathsf{Agg}[\cdot] = \frac{1}{|\mathcal{N}_h|} \sum_{(h,r,t) \in \mathcal{N}_h}[\cdot]\).

After aggregating the layers to obtain the final entity representation \(\mathbf{z}_e\) for \(e\in \mathcal{E}\), we remove the profile to recover the KG space:
\begin{align}
\mathbf{z}_e &= \sum_{l=0}^{L} \overline{\mathbf{x}}_e^{(l)}- \lambda_{p}\cdot\mathcal{M}(\mathbf{p}_e),
\label{eq:profile-strip}
\end{align}
where \(L\) is the number of graph layers and \(\overline{\mathbf{x}}_e^{(0)}\!= \overline{\mathbf{x}}_e\) for simplicity.
Although relation-aware aggregation is skipped for users (See Equation \ref{eq:profile-agg}), profile information propagates through items and auxiliary entities. 
As a result, user preferences are effectively modeled, enabling high-quality representation learning for recommendation.

\subsection{Pairwise Profile Preference Matching}
\label{sec:pairwise_matching}

The goal of this section is to align the preference between the entity and profile representations.
The motivation behind this approach is that if two profiles are highly similar, the corresponding entities should also be close in the representation space, and conversely, dissimilar profiles should lead to more distant entities.
This encourages entities with similar profiles to be placed closer together in the recommendation space.
The alignment is repeated over various randomly sampled subsets to enhance robustness.

Formally, let \( \mathcal{E} = \{e_1, \ldots, e_N\} \) denote the set of KG entities, where \(N=\left|\mathcal{E}\right|\).  
We then define the full embedding matrices as:
\begin{equation}
\mathbf{Z}_{\textsf{kg}} =
\begin{bmatrix}
\mathbf{z}_{e_1}^\top \\
\vdots \\
\mathbf{z}_{e_N}^\top
\end{bmatrix}
\in \mathbb{R}^{N \times d}, \quad
\mathbf{Z}_{\textsf{text}} =
\begin{bmatrix}
\mathcal{M}(\mathbf{p}_{e_1})^\top \\
\vdots \\
\mathcal{M}(\mathbf{p}_{e_N})^\top
\end{bmatrix}
\in \mathbb{R}^{N \times d}.
\end{equation}
For notational convenience, we assume that all row vectors are \( \ell_2 \)-normalized (\ie \(\|\mathbf{z}_e\|_2=\|\mathcal{M}(\mathbf{p}_e)\|_2=1\) for \(e\in\mathcal{E}\)).

The next step is to compute pairwise similarities.  
However, computing pairwise similarities between all entity pairs has a time complexity of \( \mathcal{O}(N^2) \), which becomes costly when \( N \) is large.  
To reduce computation and keep training stable, we perform \(R\) rounds of sampling within each training batch, where each subset of entities is selected without replacement.
Specifically, we select \( K = \left\lceil N \cdot q \right\rceil \) entities in each round, where \( q \in [0, 1] \) is the sampling ratio.
For each repetition \(r = 1, \cdots, R\), we define the sampled submatrices \(\tilde{\mathbf{Z}}^{(r)}_{\textsf{kg}}, \tilde{\mathbf{Z}}^{(r)}_{\textsf{text}}\in \mathbb{R}^{K \times d}\) as:
\begin{equation}
\tilde{\mathbf{Z}}_{\textsf{kg}}^{(r)} = \mathbf{S}^{(r)} \mathbf{Z}_{\textsf{kg}}, \quad
\tilde{\mathbf{Z}}_{\textsf{text}}^{(r)} = \mathbf{S}^{(r)} \mathbf{Z}_{\textsf{text}}
\end{equation}
where \( \mathbf{S}^{(r)} \in \{0,1\}^{K \times N} \) is an \(r_\mathsf{th}\) sampling matrix \cite{winkler2020sampling} where each row has exactly one nonzero entry, used to select \( K \) rows from each of the full embedding matrices \( \mathbf{Z}_{\textsf{kg}} \) and \( \mathbf{Z}_{\textsf{text}} \).

Finally, motivated by \cite{tung2019similarity}, we define the stochastic pairwise profile preference matching loss \(\mathcal{L}_{\text{pair}}\) as:
\begin{equation}
\mathcal{L}_{\text{pair}} = \frac{1}{R\cdot K^2} \sum_{r=1}^{R} \left\|
\tilde{\mathbf{Z}}_{\textsf{kg}}^{(r)} \left(\tilde{\mathbf{Z}}_{\textsf{kg}}^{(r)}\right)^\top -
\tilde{\mathbf{Z}}_{\textsf{text}}^{(r)} \left(\tilde{\mathbf{Z}}_{\textsf{text}}^{(r)}\right)^\top
\right\|_F^2
\label{eq:align_loss}
\end{equation}
where \( \| \cdot \|_F \) denotes the Frobenius norm.
As shown in the Equation \ref{eq:align_loss}, we adopt cosine similarity, a straightforward yet reliable measure, as the similarity function in the matching loss.
This encourages the directional alignment of profile and KG embeddings.

Note that \(\tilde{\mathbf{Z}}_{\textsf{kg}}^{(r)}\) and \(\tilde{\mathbf{Z}}_{\textsf{text}}^{(r)}\) are sampled per batch and vary across training iterations.
We omit the notation indicating batch for clarity.

\subsection{Model Training and Inference}
This section describes the overall training objective of true and how recommendation scores are computed during inference.

For the recommendation task, we predict the interaction score between a user \(u \in \mathcal{U}\) and an item \(v \in \mathcal{V}\) using the dot product of their profile-aware representations:
\begin{equation}
\hat{y}_{uv} = \mathbf{z}_u^\top \mathbf{z}_v.
\end{equation}

To train the model, we adopt the Bayesian personalized ranking (BPR) loss, following prior work \cite{yang2023kgrec, cui2025colakg, wang2019kgat}.
Specifically, the recommendation loss \(\mathcal{L}_{\text{rec}}\) is defined over the training set \(\mathcal{D} \subseteq \mathcal{U} \times \mathcal{V}^2\):
\begin{equation}
\mathcal{L}_{\text{rec}} = \sum_{(u, v^+, v^-) \in \mathcal{D}} -\log \sigma\left( \hat{y}_{uv^+} - \hat{y}_{uv^-} \right)
\end{equation}
where \(v^+\) is an item interacted with by user \(u\) (\ie \((u, r_{\mathsf{Interact}}, v^+) \in \mathcal{G}\)) and \(v^-\) is a randomly sampled item that user \(u\) has not interacted with (\ie \((u, r_{\mathsf{Interact}}, v^-) \notin \mathcal{G}\)). 
\(\sigma(\cdot)\) denotes the sigmoid function.

In addition to the recommendation objective, we apply the pairwise alignment loss described in \S\ref{sec:pairwise_matching}, which aligns the similarity structures between the graph- and profile-based embedding spaces. 
The final loss combines both objectives as:
\begin{equation}
\mathcal{L} = \mathcal{L}_{\textsf{rec}} + \lambda_{\textsf{pair}} \mathcal{L}_{\textsf{pair}},
\end{equation}
where \(\lambda_{\textsf{pair}}\) is the parameter controlling the influence of the alignment loss. 
We omit standard regularization terms for simplicity.

\section{Experimental Settings}

This section outlines the datasets, evaluation setup, and implementation details used in our experiments.

\subsubsection*{Datasets and KG construction}

We conduct experiments on three real-world datasets: Amazon Books (\textbf{Books}), Amazon Movies and TV (\textbf{Movies \& TV}), and Yelp (\textbf{Yelp}), corresponding to the recommendation domains of books, movies and TV shows, and business venues, respectively.
Even within the same dataset, prior LLM- and KG-based studies often adopt different user/item sets and item-entity connections, which require consistent entity alignment and KG construction.
Further details on dataset construction and preprocessing are provided in Appendix \ref{sec:appendix_dataset}.

\subsubsection*{Evaluation metrics}
We evaluate model performance using two widely adopted metrics: Recall and Normalized Discounted Cumulative Gain (NDCG).
Following \cite{yang2023kgrec, ren2024sslrec, cui2025colakg}, we adopt Recall@\(K\) and NDCG@\(K\) with value \(K=10,20\), and \(40\).

\subsubsection*{Baselines}
For a comprehensive evaluation, we compare SPiKE with eight state-of-the-art baseline methods categorized by whether they incorporate profiles beyond user-item interactions:
non-profiled recommenders (BPR-MF \cite{rendle2012bpr}, LightGCN \cite{he2020lightgcn}, SGL \cite{wu2021sgl}, and LightGCL \cite{cai2023lightgcl}), 
profile-aware recommenders utilizing KGs and/or LLMs (KGRec \cite{yang2023kgrec}, DiffKG \cite{jiang2024diffkg}, RLMRec \cite{ren2024rlmrec}, and CoLaKG \cite{cui2025colakg}).
Additional descriptions of the baselines are provided in Appendix \ref{sec:appendix_baseline}.

\subsubsection*{Implementation}

All experiments were conducted on a machine equipped with an AMD EPYC 7513 32‑Core CPU and an NVIDIA RTX Ada6000 GPU.
All models were implemented in PyTorch and evaluated using the unified recommendation library SSLRec \cite{ren2024sslrec} to ensure a fair comparison.
We provide additional implementation details in Appendix \ref{sec:appendix_implementation}.

\begin{table*}[t]
\centering
\caption{Main results. The highest performance is marked in \first{bold}, and the second highest in \underline{underlined}.}
\vspace{-0.7em}
\resizebox{1\textwidth}{!}{%
\setlength\tabcolsep{1.5pt}
\begin{tabular}{lrrrrrrrrrrrrrrrrrr}
\toprule
 & \multicolumn{6}{c}{\(\textbf{Books}\)} & \multicolumn{6}{c}{\(\textbf{Movies \& TV}\)} & \multicolumn{6}{c}{\(\textbf{Yelp}\)} \\
 \cmidrule(l{0.5mm}r{0.5mm}){2-7} \cmidrule(l{0.5mm}r{0.5mm}){8-13} \cmidrule(l{0.5mm}r{0.5mm}){14-19}
 & \multicolumn{3}{c}{Recall (R)\scriptsize$\uparrow$} & \multicolumn{3}{c}{NDCG (N)\scriptsize$\uparrow$} & \multicolumn{3}{c}{Recall (R)\scriptsize$\uparrow$} & \multicolumn{3}{c}{NDCG (N)\scriptsize$\uparrow$} & \multicolumn{3}{c}{Recall (R)\scriptsize$\uparrow$} & \multicolumn{3}{c}{NDCG (N)\scriptsize$\uparrow$} \\
 \cmidrule(l{0.5mm}r{0.5mm}){2-4} 
 \cmidrule(l{0.5mm}r{0.5mm}){5-7} \cmidrule(l{0.5mm}r{0.5mm}){8-10} \cmidrule(l{0.5mm}r{0.5mm}){11-13} \cmidrule(l{0.5mm}r{0.5mm}){14-16}
 \cmidrule(l{0.5mm}r{0.5mm}){17-19}
\textbf{Models} & \textbf{R@10} & \textbf{R@20} & \textbf{R@40} & \textbf{N@10} & \textbf{N@20} & \textbf{N@40} & \textbf{R@10} & \textbf{R@20} & \textbf{R@40} & \textbf{N@10} & \textbf{N@20} & \textbf{N@40} & \textbf{R@10} & \textbf{R@20} & \textbf{R@40} & \textbf{N@10} & \textbf{N@20} & \textbf{N@40} \\
 \cmidrule(l{0.5mm}r{0.5mm}){1-1} 
 \cmidrule(l{0.5mm}r{0.5mm}){2-4} 
 \cmidrule(l{0.5mm}r{0.5mm}){5-7} \cmidrule(l{0.5mm}r{0.5mm}){8-10} \cmidrule(l{0.5mm}r{0.5mm}){11-13} \cmidrule(l{0.5mm}r{0.5mm}){14-16}
 \cmidrule(l{0.5mm}r{0.5mm}){17-19}
BPR-MF & 0.1347 & 0.1779 & 0.2732 & 0.0979 & 0.1181 & 0.1396 & 0.1304 & 0.1936 & 0.2782 & 0.0934 & 0.1135 & 0.1361 & 0.0592 & 0.0960 & 0.1521 & 0.0470 & 0.0594 & 0.0762 \\
LightGCN & 0.1419 & 0.2059 & 0.2870 & 0.1038 & 0.1244 & 0.1470 & 0.1394 & 0.2073 & 0.2936 & 0.1002 & 0.1217 & 0.1449 & 0.0612 & 0.1002 & 0.1572 & 0.0487 & 0.0618 & 0.0787 \\
SGL & 0.1445 & 0.2064 & 0.2846 & 0.1079 & 0.1281 & 0.1500 & \underline{0.1433} & 0.2075 & 0.2879 & \underline{0.1057} & 0.1259 & 0.1474 & 0.0635 & 0.1018 & 0.1598 & \first{0.0512} & \first{0.0641} & \underline{0.0813} \\
LightGCL & 0.1405 & 0.2027 & 0.2772 & 0.1045 & 0.1246 & 0.1455 & 0.1420 & 0.2091 & 0.2906 & 0.1053 & \underline{0.1265} & 0.1484 & 0.0599 & 0.0977 & 0.1559 & 0.0475 & 0.0603 & 0.0776 \\
 \cmidrule(l{0.5mm}r{0.5mm}){1-1} 
 \cmidrule(l{0.5mm}r{0.5mm}){2-4} 
 \cmidrule(l{0.5mm}r{0.5mm}){5-7} \cmidrule(l{0.5mm}r{0.5mm}){8-10} \cmidrule(l{0.5mm}r{0.5mm}){11-13} \cmidrule(l{0.5mm}r{0.5mm}){14-16}
 \cmidrule(l{0.5mm}r{0.5mm}){17-19}
KGRec & \underline{0.1489} & \underline{0.2127} & \underline{0.2962} & \underline{0.1105} & \underline{0.1313} & \underline{0.1545} & 0.1393 & 0.2061 & 0.2929 & 0.1002 & 0.1215 & 0.1448 & 0.0549 & 0.0938 & 0.1499 & 0.0449 & 0.0573 & 0.0741 \\
DiffKG & 0.1413 & 0.2051 & 0.2878 & 0.1024 & 0.1231 & 0.1461 & 0.1411 & 0.2077 & 0.2973 & 0.1026 & 0.1237 & 0.1478 & 0.0623 & 0.1007 & 0.1570 & 0.0495 & 0.0623 & 0.0792 \\
RLMRec & 0.1432 & 0.2080 & 0.2868 & 0.1047 & 0.1255 & 0.1476 & 0.1431 & 0.2104 & 0.2993 & 0.1034 & 0.1247 & 0.1486 & \underline{0.0635} & 0.1006 & 0.1590 & 0.0501 & 0.0627 & 0.0799 \\
CoLaKG & 0.1428 & 0.2086 & 0.2905 & 0.1076 & 0.1287 & 0.1516 & 0.1428 & \underline{0.2106} & \underline{0.3022} & 0.1037 & 0.1254 & \underline{0.1501} & \first{0.0636} & \underline{0.1020} & \underline{0.1604} & \underline{0.0510} & \underline{0.0640} & \first{0.0814} \\
\first{SPiKE} & \first{0.1518} & \first{0.2186} & \first{0.3023} & \first{0.1120} & \first{0.1335} & \first{0.1568} 
& \first{0.1481} & \first{0.2167} & \first{0.3100} & \first{0.1068} & \first{0.1285} & \first{0.1535} 
 & 0.0631 & \first{0.1030} & \first{0.1613} & 0.0499 & 0.0633 & 0.0806 \\
\bottomrule
\end{tabular}
}
\vspace{0.2em}

\label{tb:result_main}
\end{table*}

\begin{table*}[t]
\centering
\caption{Ablation results by removing user (-u), item (-i), entity (-e) profiles, injection-removal (-r), and matching objective (-m).}
\vspace{-0.7em}
\resizebox{1\textwidth}{!}{%
\setlength\tabcolsep{1.5pt}
\begin{tabular}{lrrrrrrrrrrrrrrrrrr}
\toprule
 & \multicolumn{6}{c}{\(\textbf{Books}\)} & \multicolumn{6}{c}{\(\textbf{Movies \& TV}\)} & \multicolumn{6}{c}{\(\textbf{Yelp}\)} \\
 \cmidrule(l{0.5mm}r{0.5mm}){2-7} \cmidrule(l{0.5mm}r{0.5mm}){8-13} \cmidrule(l{0.5mm}r{0.5mm}){14-19}
 & \multicolumn{3}{c}{Recall (R)\scriptsize$\uparrow$} & \multicolumn{3}{c}{NDCG (N)\scriptsize$\uparrow$} & \multicolumn{3}{c}{Recall (R)\scriptsize$\uparrow$} & \multicolumn{3}{c}{NDCG (N)\scriptsize$\uparrow$} & \multicolumn{3}{c}{Recall (R)\scriptsize$\uparrow$} & \multicolumn{3}{c}{NDCG (N)\scriptsize$\uparrow$} \\
  \cmidrule(l{0.5mm}r{0.5mm}){2-4} 
 \cmidrule(l{0.5mm}r{0.5mm}){5-7} \cmidrule(l{0.5mm}r{0.5mm}){8-10} \cmidrule(l{0.5mm}r{0.5mm}){11-13} \cmidrule(l{0.5mm}r{0.5mm}){14-16}
 \cmidrule(l{0.5mm}r{0.5mm}){17-19}
\textbf{Models} & \textbf{R@10} & \textbf{R@20} & \textbf{R@40} & \textbf{N@10} & \textbf{N@20} & \textbf{N@40} & \textbf{R@10} & \textbf{R@20} & \textbf{R@40} & \textbf{N@10} & \textbf{N@20} & \textbf{N@40} & \textbf{R@10} & \textbf{R@20} & \textbf{R@40} & \textbf{N@10} & \textbf{N@20} & \textbf{N@40} \\
 \cmidrule(l{0.5mm}r{0.5mm}){1-1} 
 \cmidrule(l{0.5mm}r{0.5mm}){2-4} 
 \cmidrule(l{0.5mm}r{0.5mm}){5-7} \cmidrule(l{0.5mm}r{0.5mm}){8-10} \cmidrule(l{0.5mm}r{0.5mm}){11-13} \cmidrule(l{0.5mm}r{0.5mm}){14-16}
 \cmidrule(l{0.5mm}r{0.5mm}){17-19}

SPiKE-\textbf{u} & 0.1454 & 0.2107 & 0.2926 & 0.1080 & 0.1292 & 0.1521 & 0.1440 & 0.2154 & 0.3064 & 0.1054 & \underline{0.1280} & 0.1525
& 0.0630 & \underline{0.1024} & 0.1609 & 0.0497 & 0.0630 & 0.0804 \\
SPiKE-\textbf{i} & \first{0.1520} & 0.2164 & \first{0.3024} & \underline{0.1118} & \underline{0.1331} & \underline{0.1566} & 
\underline{0.1468} & \underline{0.2164} & \underline{0.3096} & \underline{0.1055} & 0.1276 & \underline{0.1526} & \first{0.0636} & 0.1022 & \underline{0.1610} & \first{0.0501} & \underline{0.0631} & \underline{0.0805} \\
SPiKE-\textbf{e} & 0.1424 & 0.2094 & 0.2958 & 0.1035 & 0.1252 & 0.1492 & 
0.1422 & 0.2131 & 0.3064 & 0.1016 & 0.1243 & 0.1491 & 0.0628 & 0.0995 & 0.1575 & 0.0496 & 0.0620 & 0.0792 \\
SPiKE-\textbf{r} & 0.0905 & 0.1369 & 0.2012 & 0.0661 & 0.0811 & 0.0990 & 0.1158 & 0.1741 & 0.2587 & 0.0825 & 0.1011 & 0.1238 & 0.0432 & 0.0718 & 0.1154 & 0.0342 & 0.0438 & 0.0569 \\
SPiKE-\textbf{m} & 0.1498 & \underline{0.2173} & 0.3012 & 0.1114 & 0.1329 & 0.1563 & 0.1438 & 0.2144 & 0.3063 & 0.1028 & 0.1253 & 0.1500 & 0.0610 & 0.0997 & 0.1595 & 0.0488 & 0.0618 & 0.0795 \\
\first{SPiKE} & \underline{0.1518} & \first{0.2186} & \underline{0.3023} & \first{0.1120} & \first{0.1335} & \first{0.1568} & \first{0.1481} & \first{0.2167} & \first{0.3100} & \first{0.1068} & \first{0.1285} & \first{0.1535} & 0.0631 & \first{0.1030} & \first{0.1613} & \underline{0.0499} & \first{0.0633} & \first{0.0806} \\
\bottomrule
\end{tabular}
}
\vspace{0.2em}

\label{tb:result_ablation}
\end{table*}

\section{Experimental Results}

In this section, we summarize the experimental results of our work.

\subsubsection*{Overall results (Table \ref{tb:result_main})}
We first compare SPiKE with the baselines in overall recommendation performance.
Overall, SPiKE outperforms both non-profiled recommenders and profile-aware recommenders in terms of Recall and NDCG.
Except for SPiKE, no profile-aware method consistently outperformed others; Knowledge-aware models performed better on \textbf{Books}, while LLM-based models achieved higher performance on \textbf{Yelp}.
\textbf{Movies \& TV} exhibits the largest performance differences across baselines, and SPiKE records its highest improvements in this domain.
Non-profile recommenders that introduced contrastive learning objectives (\ie SGL and LightGCL) did not surpass profile-aware models, but consistently delivered strong performance across all datasets.

\subsubsection*{Ablation study (Table \ref{tb:result_ablation})}
For the ablation study, we remove each key module in SPiKE to analyze their contributions.
Removing user (SPiKE-\textbf{u}), item (SPiKE-\textbf{i}), or entity (SPiKE-\textbf{e}) profiles from the KG propagation degrades performance, indicating that each profile contributes distinct information.
The small drop for item profiles (SPiKE-\textbf{i}) reflects that much of the item information is already present in the metadata, though item profiles are still needed to build user and entity profiles.
Next, the largest performance drop occurs when the removal step is omitted (SPiKE–\textbf{r}), highlighting the importance of the injection-removal process.
Finally, removing the pairwise matching loss that aligns textual profiles with KG nodes (SPiKE-\textbf{m}) also reduces performance.
Overall, all components of SPiKE contribute meaningfully to its effectiveness.

\begin{table}[t]
\centering
\vspace{0.15em}
\caption{Impact of fusion approaches on \textbf{Books} dataset.}
\vspace{-0.7em}
\resizebox{1\columnwidth}{!}{%
\setlength\tabcolsep{4pt}
\begin{tabular}{lrrrrrr}
\toprule
 & \multicolumn{3}{c}{Recall (R)\scriptsize$\uparrow$} & \multicolumn{3}{c}{NDCG (N)\scriptsize$\uparrow$} \\
\cmidrule(l{0.5mm}r{0.5mm}){2-4}
\cmidrule(l{0.5mm}r{0.5mm}){5-7}

\textbf{Fusion} & \textbf{R@10} & \textbf{R@20} & \textbf{R@40} & \textbf{N@10} & \textbf{N@20} & \textbf{N@40} \\

\cmidrule(l{0.5mm}r{0.5mm}){1-1} \cmidrule(l{0.5mm}r{0.5mm}){2-4}
\cmidrule(l{0.5mm}r{0.5mm}){5-7}
Concatenate & 0.1037 & 0.1588 & 0.2351 & 0.0732 & 0.0914 & 0.1127 \\
Attention & 0.0939 & 0.1477 & 0.2183 & 0.0657 & 0.0835 & 0.1031 \\
Mul w/o inv & \underline{0.1084} & \underline{0.1613} & \underline{0.2287} & \underline{0.0781} & \underline{0.0952} & \underline{0.1139} \\
Mul w/ inv & 0.0397 & 0.0636 & 0.1013 & 0.0277 & 0.0356 & 0.0462 \\
Add w/o inv & 0.0905 & 0.1369 & 0.2012 & 0.0661 & 0.0811 & 0.0990 \\
\first{Add w/ inv} & \first{0.1518} & \first{0.2186} & \first{0.3023} & \first{0.1120} & \first{0.1335} & \first{0.1568} \\
\bottomrule
\end{tabular}
}
\\[0.3em]
\leftline{\small{\;\;* \textbf{w/o inv}: additive or multiplicative fusion without applying inverse.}\;}
\vspace{-0.1em}
\leftline{\small{\;\;* \textbf{w/ inv}: additive or multiplicative fusion with inverse applied.}\;}

\label{tb:result_integration}
\end{table}

\subsubsection*{Impact of fusion approaches (Table \ref{tb:result_integration})}
We further compare the profile-aware KG aggregation design with several fusion alternatives:
concatenation (Concatenate), 
attentive fusion (Attention), 
multiplication only (Mul w/o inv), 
multiplication with inverse (Mul w/ inv),
addition only (Add w/o inv), and addition with inverse (Add w/ inv).
As shown in Table~\ref{tb:result_integration}, concatenation and attention perform better than addition only but remain limited due to the lack of an inverse operation.
Multiplication-based variants are invertible and outperform their non-inverse counterparts, but applying the inverse step causes performance degradation because near-zero divisions introduce numerical instability.
Only the additive inverse remains stable and further improves performance.
Therefore, SPiKE adopts additive fusion with inverse (Add w/ inv), which provides the best performance while remaining simple and invertible.

\subsubsection*{Impact of profile generation methods (Table \ref{tb:result_profiling_model})}
Next, we examine how different profile generation methods affect performance.
We compare four different settings: no profiles (None), a textual template (Template), template profiles refined by an open-source LLM (Qwen3-4B), and a closed-source LLM (GPT-4o). 
The improvement from None to Template shows that even adding textual cues offers benefits beyond structural information alone.
Refinements from both Qwen3-4B and GPT-4o further enhance performance, with GPT-4o yielding the larger gain.
These results suggest that our profiling method remains effective even when using raw template-based profiles and generalizes well across both open- and closed-source LLMs.
Template examples are listed in Appendix \ref{sec:appendix_template}.

\subsubsection*{Automated profile quality evaluations (Table \ref{tb:result_profile_quality})}
We assess the quality of generated profiles against three representative user profiling methods (\ie RLMRec, CoLaKG, and SPiKE).
For automated evaluations, we adopt \texttt{GPT-5-nano} as the evaluator and measure \textbf{Persuasiveness}, \textbf{Relevance}, and \textbf{Faithfulness} on a 1-5 Likert scale \cite{guo2023towards, ramos2024transparent, tsai2024leveraging}.
The results show that our method maintains strong performance across all three criteria, producing high-quality profiles.
Viewed together with Table \ref{tb:result_profiling_model}, these results show that high-quality profiles also improve downstream accuracy, indicating that higher-quality profiles translate into better recommendation performance.
We provide in Appendix \ref{sec:appendix_criteria} the detailed descriptions and the justification for selecting these criteria.

\begin{table}[t]
\centering
\caption{Impact of profiling methods on \textbf{Books} dataset.}
\vspace{-0.7em}
\resizebox{0.95\columnwidth}{!}{%
\setlength\tabcolsep{4pt}
\begin{tabular}{lrrrrrr}
\toprule
 & \multicolumn{3}{c}{Recall (R)\scriptsize$\uparrow$} & \multicolumn{3}{c}{NDCG (N)\scriptsize$\uparrow$} \\
\cmidrule(l{0.5mm}r{0.5mm}){2-4}
\cmidrule(l{0.5mm}r{0.5mm}){5-7}

\textbf{Methods} & \textbf{R@10} & \textbf{R@20} & \textbf{R@40} & \textbf{N@10} & \textbf{N@20} & \textbf{N@40} \\

\cmidrule(l{0.5mm}r{0.5mm}){1-1} \cmidrule(l{0.5mm}r{0.5mm}){2-4}
\cmidrule(l{0.5mm}r{0.5mm}){5-7}
None & 0.1489 & 0.2127 & 0.2962 & 0.1105 & 0.1313 & 0.1545 \\
Template & 0.1491 & 0.2156 & 0.3001 & 0.1099 & 0.1320 & 0.1554 \\
Qwen3-4B & \underline{0.1508} & \underline{0.2176} & \underline{0.3023} & \underline{0.1107} & \underline{0.1324} & \underline{0.1559} \\
\first{GPT-4o} & \first{0.1518} & \first{0.2186} & \first{0.3023} & \first{0.1120} & \first{0.1335} & \first{0.1568} \\
\bottomrule
\end{tabular}
}
\vspace{0.2em}
\label{tb:result_profiling_model}
\end{table}

\begin{table}[t]
\centering
\caption{Automated profile quality evaluations (range 1 to 5) across different profiling strategies on \textbf{Books} dataset.}
\vspace{-0.7em}
\resizebox{0.92\columnwidth}{!}{%
\setlength\tabcolsep{4pt}
\begin{tabular}{lccc}
\toprule
\textbf{Strategies} & \textbf{Persuasiveness}\scriptsize$\uparrow$ & \textbf{Relevance}\scriptsize$\uparrow$ & \textbf{Faithfulness}\scriptsize$\uparrow$ \\
\cmidrule(l{0.5mm}r{0.5mm}){1-1} \cmidrule(l{0.5mm}r{0.5mm}){2-4}
RLMRec & 3.706 & 3.826 & 3.679 \\
CoLaKG & 3.638 & 3.825 & 3.665 \\
\first{SPiKE} & \first{3.712} & \first{3.838} & \first{3.699} \\
\bottomrule
\end{tabular}
}
\label{tb:result_profile_quality}
\end{table}

\begin{figure}
    \centering
    \vspace{-0.4em}
    \includegraphics[width=0.94\columnwidth]{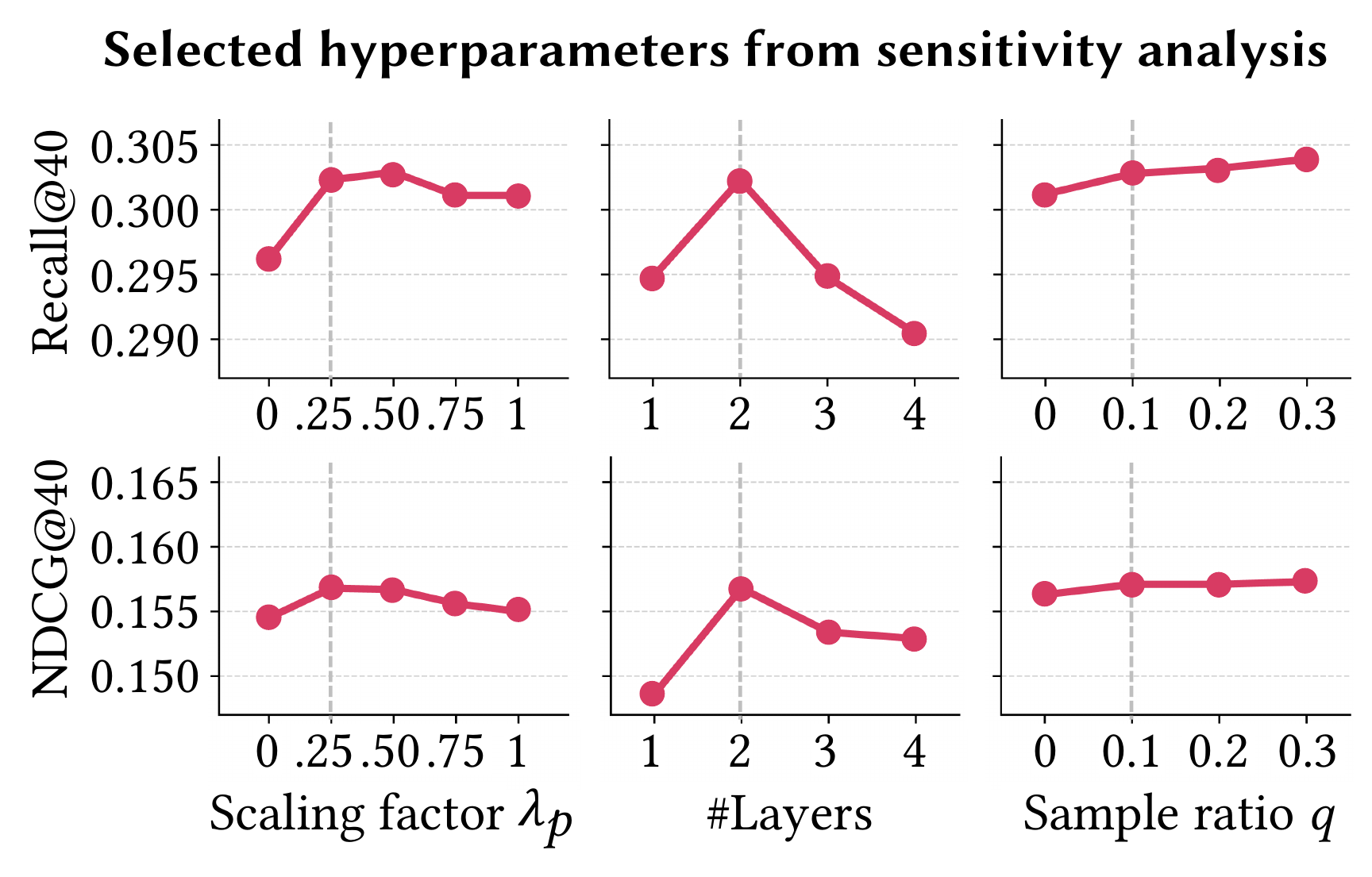}
    \vspace{-1.2em}
    \caption{Sensitivity analysis of key SPiKE hyperparameters on Books dataset, with final selections marked by \textcolor[HTML]{777777}{gray vertical dashed lines}. All choices are made by considering performance across all @$K$ metrics and efficiency trade-offs.}
    \Description[]{Sensitivity analysis of key SPiKE hyperparameters on Books dataset, with final selections marked by \textcolor[HTML]{777777}{gray vertical dashed lines}. All choices are made by considering performance across all @$K$ metrics and efficiency trade-offs.}
    \label{fig:result_hyperparameter}
\end{figure}

\subsubsection*{Hyperparameter study (Figure \ref{fig:result_hyperparameter})}
We analyze the hyperparameter sensitivity of three factors:
the scaling factor $\lambda_p$ used in entity profiling (\S\ref{sec:entity_profiling}), 
the number of KG propagation layers (\S\ref{sec:kg_aggregation}), and 
the sampling ratio $q$ controlling efficient pairwise matching (\S\ref{sec:pairwise_matching}). 
Results are reported at Recall@40 and NDCG@40 in Figure \ref{fig:result_hyperparameter}, but all choices are selected based on performance across $K\in\left\{10, 20, 40\right\}$.
For the scaling factor $\lambda_p$, performance is highest when $\lambda_p$ lies between 0.25 and 0.5, with only minor differences within this range.
For the number of aggregation layers, two layers achieve the highest accuracy, while deeper propagation reduces accuracy due to over-smoothing in KG embeddings.
For the sampling ratio $q$, performance grows only slightly as $q$ increases.
Because computational cost scales quadratically with $q$, a small ratio (\ie $q=0.1$) is sufficient.
These configurations are used for SPiKE on \textbf{Books} dataset.

\begin{figure}
    \centering
    \vspace{-0.3em}
    \includegraphics[width=0.94\columnwidth]{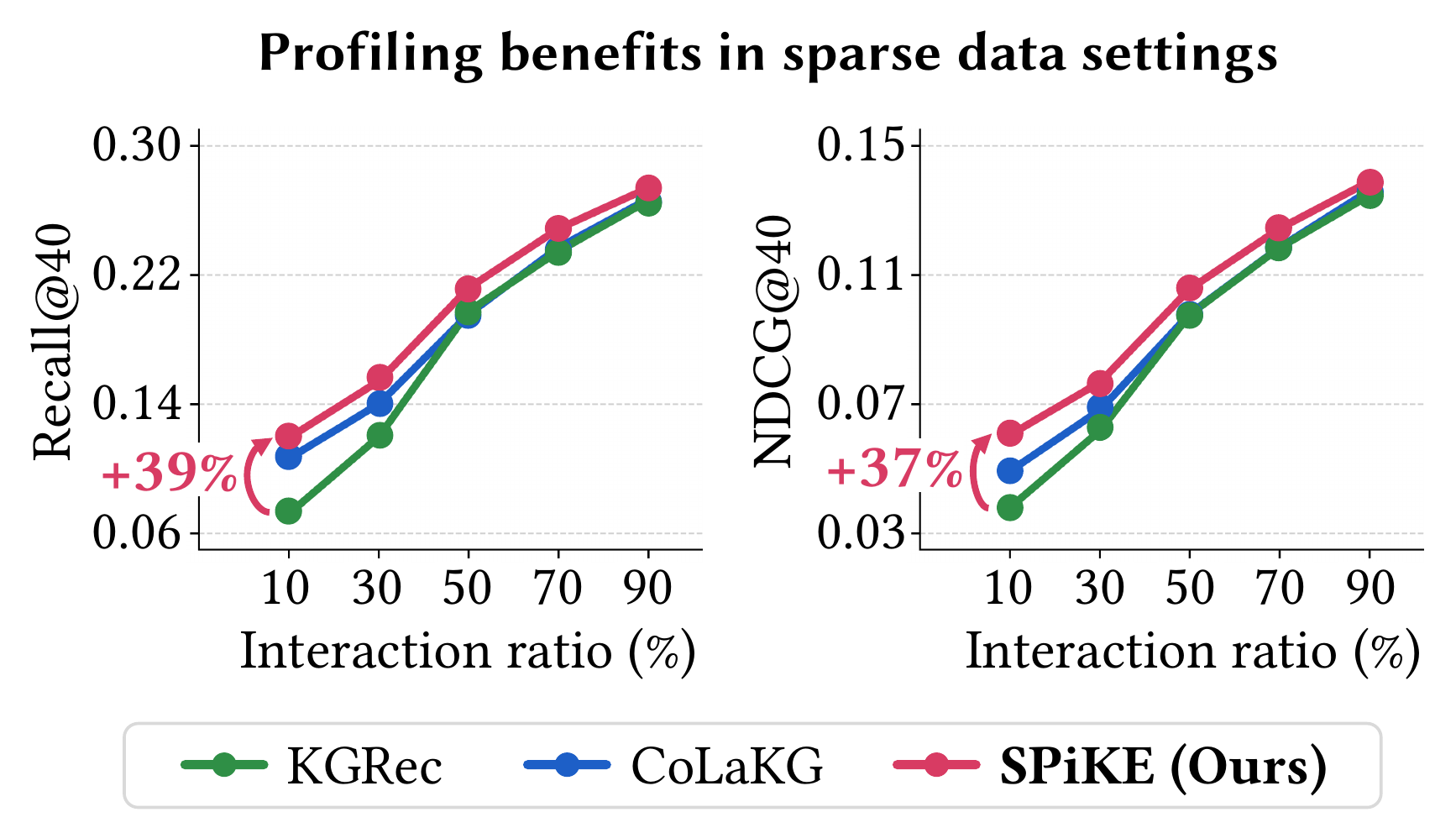}
    \vspace{-0.8em}
    \caption{Performance comparison under different graph-density settings on Books dataset, controlled by varying the interaction ratio. Each user keeps at least one interaction.}
    \Description[]{Performance comparison under different graph-density settings on Books dataset, controlled by varying the interaction ratio. Each user keeps at least one interaction.}
    \vspace{-0.5em}
    \label{fig:result_coldstart}
\end{figure}

\subsubsection*{Sparse data settings (Figure \ref{fig:result_coldstart})}
We evaluate model behavior under different graph structures by varying the interaction ratio, covering scenarios from cold/sparse (10\%) to warm/dense settings (90\%).
At the lowest interaction ratios, SPiKE exhibits a noticeably larger margin over the baselines (especially KGRec), demonstrating strong robustness under cold-start conditions.
This result suggests that semantic profiles provide additional guidance when graph connectivity is limited.
Although all models improve with more interactions, SPiKE maintains its lead, showing that its advantage extends beyond sparse settings.
Overall, SPiKE maintains consistent performance across different sparsity conditions.

\begin{figure}
    \centering
    \vspace{-0.2em}
    \includegraphics[width=0.82\columnwidth]{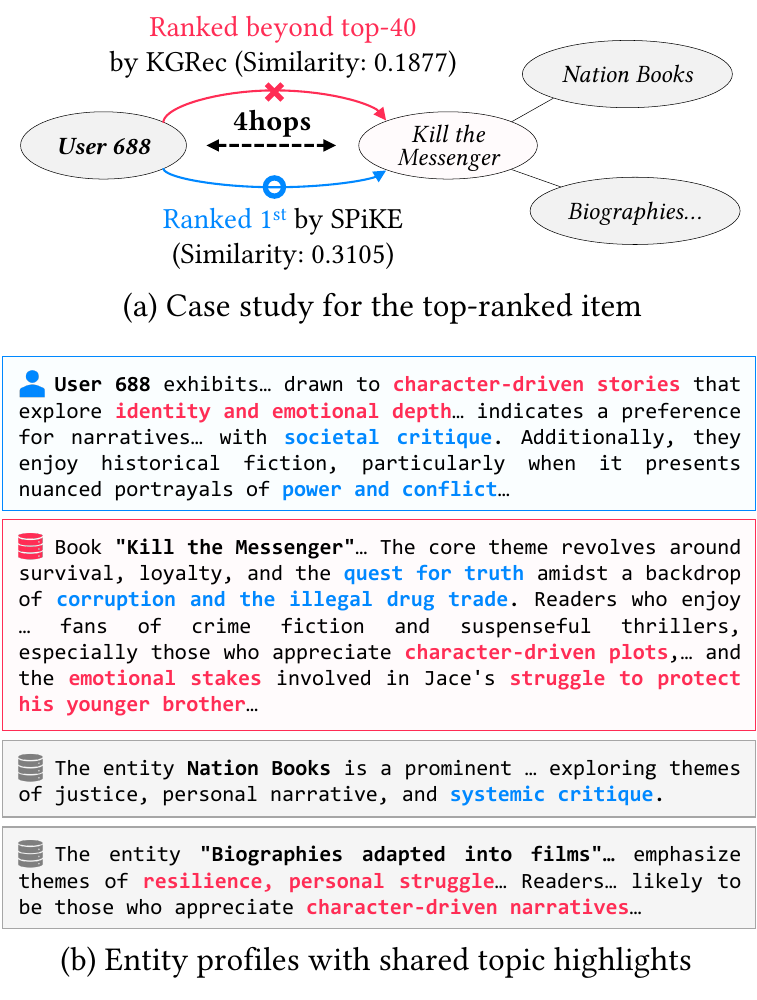}
    \vspace{-1.5em}
    \caption{Case studies on Books dataset.}
    \Description[]{Case studies on Books dataset.}
    \label{fig:case}
    \vspace{-0.7em}
\end{figure}

\subsubsection*{Case Study (Figure \ref{fig:case})}
We present a case study to visually examine the influence of profile.
To avoid selection bias, we choose the user for whom KGRec achieves the highest recommendation accuracy.
For this user, item \textsf{Kill the Messenger} ranks first under SPiKE but falls outside the top-40 under KGRec. 
The item lies beyond a 3-hop neighborhood of the user node, so KGRec cannot reach it structurally (See Figure \ref{fig:case}(a)).
Figure \ref{fig:case}(b) shows that the user description, the item, and related entities share topics (highlighted in the same color). 
This example shows that textual profiles supply complementary semantics, allowing SPiKE to retrieve relevant items unreachable by structure alone.

\section{Discussions}
We discuss the implications of the experimental results and potential limitations of our approach.

\subsubsection*{Impact of using knowledge-aware reasoning}
Table \ref{tb:result_main} shows that the benefit of adding KG signals depends heavily on KG quality.
The \textbf{Books} dataset relies on a large, well-structured graph from \cite{wang2024amazonkg}, and knowledge-aware models gain significantly.  
By contrast, the \textbf{Yelp} KG, constructed from metadata, is smaller and less structured, which limits the effectiveness of knowledge-aware models.
This suggests that incorporating KG signals enhances recommendation performance by leveraging structural reasoning, but additional strategies are needed for weaker KGs.

\subsubsection*{Impact of using LLM-generated profiles}
As shown in Table \ref{tb:result_main}, LLM-generated profiles can provide semantic clues about user preferences that are often unavailable from the KG alone.
Specifically, in the \textbf{Yelp} setting, LLM-based methods outperform knowledge-aware methods.
This is due to the LLMs’ before construct profiles that filter noise and inject missing knowledge from various knowledge bases.
However, relying solely on LLM-generated profiles leads to degraded performance in large-scale KG settings (\eg \textbf{Books}), since LLMs cannot utilize structural signals such as multi-hop relations.

\subsubsection*{Importance of how KG and LLM are combined}
The strengths of KG and LLM are complementary, but how they are combined plays a critical role in recommendation performance.
While proper integration can improve effectiveness, a poorly designed strategy may limit the benefits of both.
CoLaKG tries to incorporate both sources but struggles to utilize the global KG effectively, leading to performance degradation as KG size grows (\ie limited profile reach).
SPiKE only shows consistent performance across all datasets by combining LLM-based profiling and KG-based propagation.

\subsubsection*{Profiling beyond utilizing factual knowledge}
The profiling process is not primarily intended to extract factual knowledge about items.
Rather, it summarizes user preferences scattered in the structured relationships (\eg ``This item tends to be preferred by users with specific interests.'').
For most items, summarizing factual descriptions is less meaningful because the LLM may already know them.
Instead, preference signals attached to users and entities contribute most to profile quality (see Table \ref{tb:result_ablation}).
Although the LLM has no direct access to raw user histories, it can still construct meaningful profiles by relying on preference information encoded in the KG.


\subsubsection*{Scalability and efficiency}
Hybrid models combining KGs and LLMs generally require higher computational overhead than KG-only or LLM-only methods.
As discussed in \S\ref{sec:02_overview}, SPiKE mitigates this issue by excluding LLMs from the training loop.
LLMs are used only once for profile generation, which is a one-time pre-computation and does not affect subsequent training.
The training complexity then grows linearly with the number of entities, driven mainly by profiling-aware aggregation and the sampling ratio used for pairwise matching.
In practice, SPiKE trains at roughly twice the speed of KGRec, showing that it remains scalable and computationally efficient even with LLM-generated profiles.

\subsubsection*{Further learning strategy}
We design SPiKE to be simple yet effective, combining LLM-generated profiles with KG-based propagation while avoiding specialized training tricks.
This design helps highlight the true performance gains from each component.
Despite its minimal design, SPiKE outperforms existing models that employ complex training techniques.
This highlights its potential for further improvement.
As shown in Table \ref{tb:result_main}, even general recommenders using contrastive learning achieve strong performance without additional external knowledge.
Incorporating such learning strategies into SPiKE could further enhance its effectiveness.

\section{Conclusion}
This paper revisits profiling in recommender systems through four key dimensions: knowledge base, preference indicator, impact range, and subject.
We then propose SPiKE, a model that follows design rationale via semantic profiles enriched by LLMs and propagated through a KG. 
SPiKE consists of three components: entity profile generation, constructing high-quality textual profiles for all KG entities; profile-aware KG aggregation, jointly propagating structural and semantic signals; and pairwise profile preference matching, aligning profile similarities in the embedding space.
This design enables broader preference modeling and yields consistent performance gains across diverse benchmarks. 
Our findings highlight the importance of both profile construction and propagation strategy when integrating LLMs with KGs for recommendation.

\begin{acks}
This work was supported by the National Research Foundation of Korea (NRF) grant funded by the Korea government (MSIT) (No. RS-2025-24683400), Institute of Information \& communications Technology Planning \& Evaluation (IITP) grants funded by the Korea government (MSIT) (No. 2022-0-00448/RS-2022-II220448, Deep Total Recall: Continual Learning for Human-Like Recall of Artificial Neural Networks), and  INHA UNIVERSITY Research Grant.
\end{acks}

\bibliographystyle{ACM-Reference-Format}
\bibliography{00_main}

\appendix
\section{Dataset Details \label{sec:appendix_dataset}}

\begin{table}[t]

\centering
\setlength\tabcolsep{4pt}
\caption{Dataset statistics used in our experiments.}
\vspace{-0.7em}
\resizebox{0.92\columnwidth}{!}{%
\begin{tabular}{lcccc}
    \toprule 
    \textbf{Statistics} & \textbf{Books} & \textbf{Movies \& TV} & \textbf{Yelp} \\
    \cmidrule(l{0.5mm}r{0.5mm}){1-1} \cmidrule(l{0.5mm}r{0.5mm}){2-4}
    \# Users & 14,946 & 17,675 & 18,459 \\
    \# Items & 8,696 & 4,896 & 15,738 \\
    \# User-item interactions & 118,491 & 130,657 & 283,183 \\
    \cmidrule(l{0.5mm}r{0.5mm}){1-1} \cmidrule(l{0.5mm}r{0.5mm}){2-4}
    \# Auxiliary entities & 23,687 & 34,342 & 1,058 \\
    \# Relations & 20 & 20 & 3 \\
    \# Triplets w/o interactions  & 69,573 & 106,763 & 101,003 \\
    \bottomrule
\end{tabular}
\label{tb:dataset}
}
\vspace{-0.5em}
\label{tb:metadata}
\end{table}

\begin{table*}[t]
\centering
\caption{Examples of KG relation (1-hop and 2-hop) templates used to generate natural language descriptions on Books dataset.}
\vspace{-0.7em}
\resizebox{\textwidth}{!}{%
\begin{tabular}{p{0.15\linewidth} p{0.38\linewidth} p{0.47\linewidth}}
\toprule
\textbf{KG relation} & \textbf{1-hop textual template} & \textbf{2-hop textual template} \\
\midrule

\texttt{subject} &
\texttt{[ITEM]} is about \texttt{[ENTITY]}. &
\texttt{[ENTITY]} also appears as a subject in items such as \texttt{[ITEMS]}. \\

\texttt{author} &
\texttt{[ITEM]} lists \texttt{[ENTITY]} as its author. &
\texttt{[ENTITY]} also wrote other works like \texttt{[ITEMS]}. \\

\texttt{literaryGenre} &
\texttt{[ITEM]} belongs to the literary genre \texttt{[ENTITY]}. &
The literary genre `\texttt{[ENTITY]}' also includes items such as \texttt{[ITEMS]}. \\

\texttt{subsequentWork} &
\texttt{[ITEM]} is followed by \texttt{[ENTITY]}. &
\texttt{[ENTITY]} is followed by items such as \texttt{[ITEMS]}. \\

\texttt{previousWork} &
\texttt{[ITEM]} comes after \texttt{[ENTITY]}. &
\texttt{[ENTITY]} is preceded by works like \texttt{[ITEMS]}. \\

\texttt{series} &
\texttt{[ITEM]} is part of the series called \texttt{[ENTITY]}. &
The series `\texttt{[ENTITY]}' also includes items such as \texttt{[ITEMS]}. \\

\texttt{director} &
\texttt{[ITEM]} was directed by \texttt{[ENTITY]}. &
\texttt{[ENTITY]} also directed other items like \texttt{[ITEMS]}. \\

\texttt{musicComposer} &
\texttt{[ITEM]} features music composed by \texttt{[ENTITY]}. &
\texttt{[ENTITY]} also composed music for works such as \texttt{[ITEMS]}. \\

\texttt{producer} &
\texttt{[ITEM]} was produced by \texttt{[ENTITY]}. &
\texttt{[ENTITY]} also produced items like \texttt{[ITEMS]}. \\

\texttt{starring} &
\texttt{[ITEM]} stars \texttt{[ENTITY]}. &
\texttt{[ENTITY]} also starred in items such as \texttt{[ITEMS]}. \\

\texttt{writer} &
\texttt{[ITEM]} was written by \texttt{[ENTITY]}. &
\texttt{[ENTITY]} also contributed to writing items like \texttt{[ITEMS]}. \\

\texttt{genre} &
\texttt{[ITEM]} is categorized under the genre \texttt{[ENTITY]}. &
The genre `\texttt{[ENTITY]}' also includes items such as \texttt{[ITEMS]}. \\

\texttt{composer} &
\texttt{[ITEM]} includes compositions by \texttt{[ENTITY]}. &
\texttt{[ENTITY]} also composed other items like \texttt{[ITEMS]}. \\

\texttt{creator} &
\texttt{[ITEM]} was created by \texttt{[ENTITY]}. &
\texttt{[ENTITY]} also created works such as \texttt{[ITEMS]}. \\

\texttt{executiveProducer} &
\texttt{[ITEM]} had \texttt{[ENTITY]} as executive producer. &
\texttt{[ENTITY]} also served as executive producer for items like \texttt{[ITEMS]}. \\

\texttt{notableWork} &
\texttt{[ITEM]} is best known for \texttt{[ENTITY]}. &
\texttt{[ENTITY]} is also notable in works such as \texttt{[ITEMS]}. \\

\texttt{award} &
\texttt{[ITEM]} received the award titled \texttt{[ENTITY]}. &
Items that received the `\texttt{[ENTITY]}' award also include \texttt{[ITEMS]}. \\

\texttt{portrayer} &
\texttt{[ITEM]} features a character portrayed by \texttt{[ENTITY]}. &
\texttt{[ENTITY]} also portrayed characters in items such as \texttt{[ITEMS]}. \\

\texttt{album} &
\texttt{[ITEM]} is included in the album \texttt{[ENTITY]}. &
The album `\texttt{[ENTITY]}' also contains items such as \texttt{[ITEMS]}. \\

\texttt{artist} &
\texttt{[ITEM]} features a performance by \texttt{[ENTITY]}. &
\texttt{[ENTITY]} also performed in items like \texttt{[ITEMS]}. \\

\bottomrule
\end{tabular}
}
\vspace{0.35em}
\label{tb:template}
\end{table*}

\begin{table*}[t]
\centering
\vspace{-0.2em}
\caption{Evaluation criteria for profile quality on Books dataset.}
\vspace{-0.7em}
\resizebox{1\textwidth}{!}{%
\begin{tabular}{p{0.16\linewidth} p{0.84\linewidth}}
\toprule
\textbf{Criterion} & \textbf{Description} \\
\cmidrule(l{0.5mm}r{0.5mm}){1-2}

\textbf{Persuasiveness \cite{guo2023towards}} &
Measures how convincingly the profile communicates the user's reading preferences and underlying motivations. 
A high score requires not only clear and logical reasoning but also meaningful subjective signals, such as sentiments, opinions, or preference rationales, that explain why the user engages with certain types of books. 
A low score means the profile is generic, impersonal, or lacks subjective insight into the user's preferences. 
\\





\cmidrule(l{0.5mm}r{0.5mm}){1-2}
\textbf{Relevance \cite{ramos2024transparent}} &
Measures how well the profile reflects the actual reading preferences shown in the user’s past interactions. 
A low score means the described preferences deviate significantly from the books the user interacted with or misrepresent their interests. 
A high score means the profile accurately captures dominant themes, genres, and patterns present in the user’s history, closely aligning with their demonstrated behavior. \\
\cmidrule(l{0.5mm}r{0.5mm}){1-2}
\textbf{Faithfulness \cite{tsai2024leveraging}} &
Measures how factually grounded the profile is in the provided interaction history, without hallucinations or contradictions. 
A low score means the profile includes unsupported claims, fabricated preferences, or statements that contradict the given evidence. 
A high score means all claims can be directly supported or reasonably inferred from the interaction history, with no factual errors or invented information. \\

\bottomrule
\end{tabular}
}
\vspace{0.44em}
\label{tb:criteria}
\end{table*}

Table \ref{tb:dataset} summarizes their basic statistics of \textbf{Books}, \textbf{Movies \& TV}, and \textbf{Yelp}.
We find prior LLM- and KG-based studies often report inconsistent statistics even on the same dataset. 
To ensure fair comparison, we follow the Amazon-KG \cite{wang2024amazonkg} structure for \textbf{Books} and \textbf{Movies \& TV}, and adopt the sampling method used in \cite{ren2024rlmrec} for \textbf{Yelp} using internal business metadata for item linking.
We then construct users and entities after aligning the items.
We follow prior work for data preprocessing \cite{ren2024rlmrec, cui2025colakg, yang2023kgrec}, applying consistent user/item $K$-core filtering, a 7:1:2 split for training, validation, and test sets, and one-to-one negative sampling per positive interaction.

\section{Baseline Details \label{sec:appendix_baseline}}
We compare SPiKE with eight state-of-the-art baseline methods categorized by non-profiled recommenders (BPR-MF \cite{rendle2012bpr}, LightGCN \cite{he2020lightgcn}, SGL \cite{wu2021sgl}, and LightGCL \cite{cai2023lightgcl}) and profile-aware recommenders (KGRec \cite{yang2023kgrec}, DiffKG \cite{jiang2024diffkg}, RLMRec \cite{ren2024rlmrec}, and CoLaKG \cite{cui2025colakg}).

Descriptions of the baselines are provided below:

\begin{itemize}[noitemsep, leftmargin=1.1em]
    \item \textbf{BPR-MF} \cite{rendle2012bpr} is a matrix factorization model optimized with the BPR loss to capture user preferences from implicit feedback.
    
    \item \textbf{LightGCN} \cite{he2020lightgcn} is a lightweight GNN that simplifies activations and feature transformations while retaining effectiveness.

    \item \textbf{SGL} \cite{wu2021sgl} enhances GNN-based recommender by leveraging the contrastive learning paradigm to improve node representations.

    \item \textbf{LightGCL}~\cite{cai2023lightgcl} applies SVD-guided contrastive learning to preserve semantics and align user-specific and global signals.

    \item \textbf{KGRec}~\cite{yang2023kgrec} uses rationale scores from attentive knowledge rationalization to guide generative and contrastive self-supervised learning in knowledge-aware recommendation.

    \item \textbf{DiffKG}~\cite{jiang2024diffkg} integrates a diffusion model with KG-specific augmentation to filter noisy item-entity connections and learn robust knowledge-aware representations.

    \item \textbf{RLMRec}~\cite{ren2024rlmrec} enhances recommender systems by aligning LLM-generated semantic profiles with collaborative signals through a cross-view alignment framework.

    \item \textbf{CoLaKG}~\cite{cui2025colakg} enables LLMs to understand KGs through item centered subgraph and user-item interaction graph, while leveraging retrieval-based similarity for global KG utilization.
\end{itemize}

\section{Implementation Details \label{sec:appendix_implementation}}

Using unified recommendation library SSLRec \cite{ren2024sslrec}, training was performed for 500 epochs with early stopping, using the Adam optimizer and an embedding size of 64.
For knowledge-aware models, the batch size was set to 1024, while other models used a batch size of 4096. 
These settings are consistent with prior work \cite{ren2024rlmrec, yang2023kgrec, jiang2024diffkg}.
We search the learning rate within the range of $1\times10^{-5}$ to $5\times10^{-4}$.
Also, following prior work \cite{ren2024rlmrec, cui2025colakg}, model-agnostic methods (\ie SGL, LightGCL, RLMRec and CoLaKG) adopt LightGCN as the backbone for fair comparison.

We leverage the \texttt{GPT-4o-mini-2024-07-18}\footnote{https://arxiv.org/abs/2410.21276} model to generate entity profiles, as it offers a good balance between performance and cost.
We also adopt \texttt{Qwen3-4B}\footnote{https://arxiv.org/abs/2505.09388} as an open-source model to verify the generalizability of LLM-based profiling, choosing it for its competitive performance against closed-source models.
For evaluating profile quality, we additionally employ \texttt{GPT-5-nano}\footnote{https://cdn.openai.com/gpt-5-system-card.pdf} as a judge model with medium reasoning effort.
For the text embedding model, we use the pre-trained \texttt{sup-simcse-roberta-large} model \cite{gao2021simcse}, following prior work \cite{cui2025colakg}.

\section{Template Profiles \label{sec:appendix_template}}
As shown in Table \ref{tb:template}, template profiles are constructed by converting 1-hop and 2-hop KG relations into predefined textual patterns.
We additionally use reversed 1-hop templates for entity-side descriptions (\eg ``\texttt{[ENTITY]} is the subject of \texttt{[ITEM]}''), although these examples are omitted here due to space constraints.
Using 2-hop connections, these templates summarize user-item interactions and item-entity relations, enabling profile generation for users, items, and entities.
They also capture collaborative filtering patterns, such as users connected through shared items or items linked through shared users.
When these templates are propagated via KG, a two-layer propagation extends the receptive field to four hops.

Note that these profiles still enhance SPiKE when propagated through the KG.
As shown in Table \ref{tb:result_profile_quality}, replacing LLM-generated profiles with templates (\ie None \textit{\textsf{vs.}} Template) still outperforms existing aggregation methods.
This indicates that simply incorporating textual profiles into KG aggregation improves performance, and refinement by an LLM further amplifies the effect.
In summary, SPiKE’s advantage stems not from LLM capabilities alone, but in the effective integration of textual profiles into KG-based propagation.

\section{Evaluation Criteria \label{sec:appendix_criteria}}
We evaluate profile quality based on persuasiveness, relevance, and faithfulness, as summarized in Table \ref{tb:criteria}.
These criteria capture complementary aspects of profile quality, covering subjective reasoning (\textbf{Persuasiveness}), behavioral alignment (\textbf{Relevance}), and factual grounding (\textbf{Faithfulness}) \cite{guo2023towards, ramos2024transparent, tsai2024leveraging}.
The evaluation prompts included detailed scoring guidelines for all levels (1 to 5). 


\end{document}